\documentclass[allclo]{FBSart}
\usepackage{amsfonts}
\usepackage{amssymb}
\usepackage{graphicx}

\newcommand{\beqn}{\begin{equation}}
\newcommand{\eeqn}{\end{equation}}
\newcommand{\bea}{\begin{eqnarray}}
\newcommand{\eea}{\end{eqnarray}}
\newcommand{\Hzero}{T_{\rm rel}}
\newcommand{\flow}{s}

\title{Similarity Renormalization Group for Few-Body Systems}
\author{R. J. Furnstahl\thanks{\textit{E-mail address:} 
furnstahl.1@osu.edu}}
\institute{Department of Physics, The Ohio State University, 
Columbus, Ohio,
USA}

\runningauthor{R.\,J.\,Furnstahl}
\runningtitle{SRG for Few-Body Systems}
\begin{document}

\maketitle
\begin{abstract}
Internucleon interactions evolved via flow equations yield
soft potentials that lead to
rapid variational convergence in few-body systems.
\end{abstract}


The Similarity Renormalization Group
(SRG)~\cite{Glazek:1993rc,Wegner:1994}  provides a compelling
method for evolving internucleon forces to softer 
forms by decoupling low- from high-momentum matrix 
elements~\cite{Bogner:2006srg,Bogner:2007srg}. 
A series of unitary transformations parameterized by $\flow$
(or $\lambda \equiv s^{-1/4}$) is
implemented through a flow equation:
\beqn
H_\flow = U_\flow H U^\dagger_\flow \equiv \Hzero + V_\flow 
 \quad \Longrightarrow \quad
 \frac{dH_\flow}{d\flow} = [ [G_\flow, H_\flow], H_\flow] \;,
\label{eq:Hflow}
\eeqn
where $\Hzero$ is the relative kinetic energy.
Applications to nuclear physics to date in a partial-wave
momentum basis have used 
$G_\flow = T_{\rm rel}$~\cite{Bogner:2006srg}, 
so the flow equation for each matrix element is
(with $\epsilon_k \equiv \langle k | \Hzero | k  \rangle = \hbar^2
k^2/m$)  
\beqn
  \frac{d}{ds} \langle k|V_\flow|k'\rangle =
     -(\epsilon_k - \epsilon_{k'})^2 \langle k | V_s | k' \rangle
    +    \sum_q (\epsilon_k + \epsilon_{k'} - 2 \epsilon_q) 
      \langle k|V_\flow|q\rangle
      \langle q|V_\flow|k'\rangle
      \;.  \label{eq:mecommutatorT}
\eeqn
The flow of 
off-diagonal matrix elements is dominated by the first term,
which drives them rapidly to zero.
This partially diagonalizes the momentum-space potential,
leading to 
decoupling~\cite{Bogner:2007srg}.
Pictures showing different initial NN potentials evolving to band-diagonal
form can be viewed at the
SRG website~\cite{srgwebsite}.

\begin{figure}[hbt]
 \includegraphics*[width=2.5in]{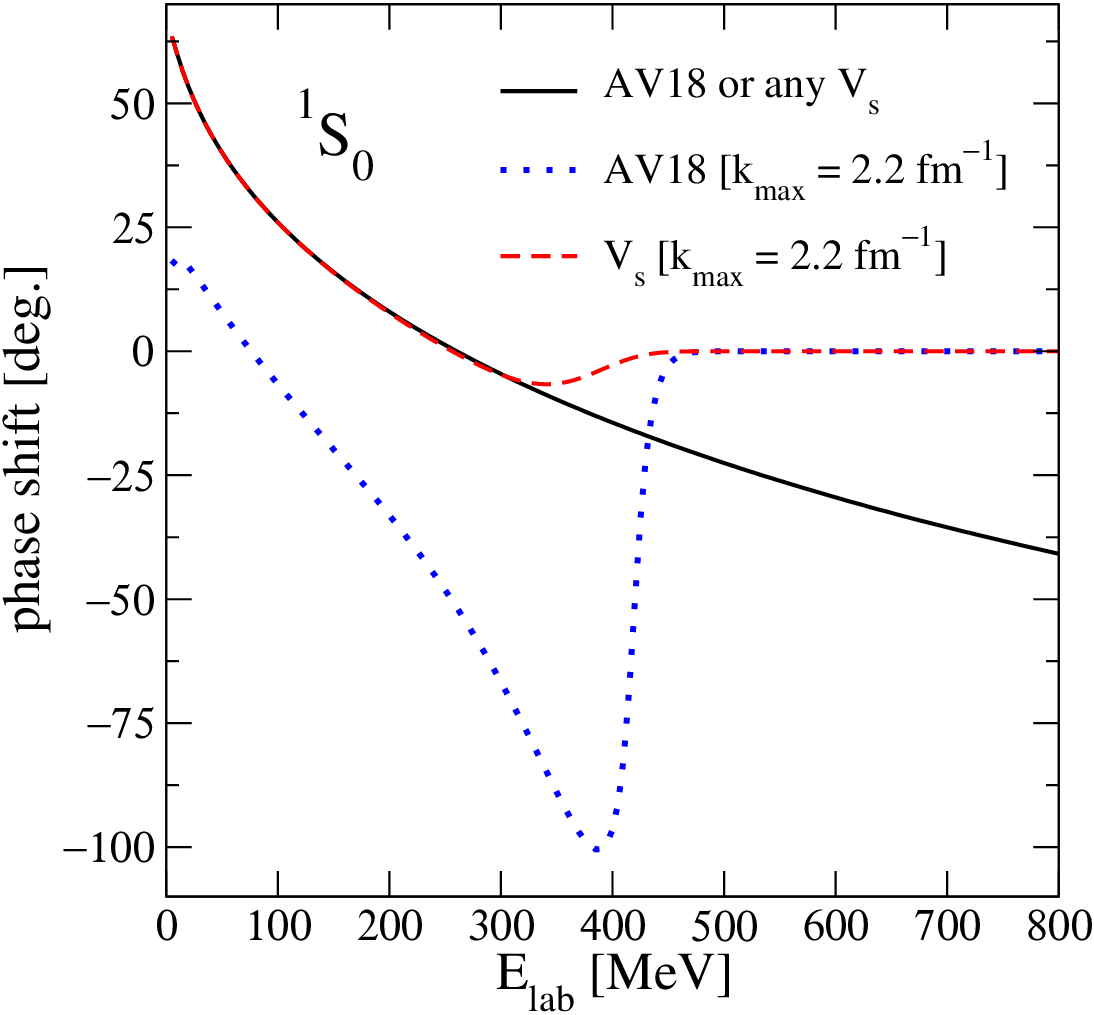}
 \hfill 
 \includegraphics*[width=2.36in]{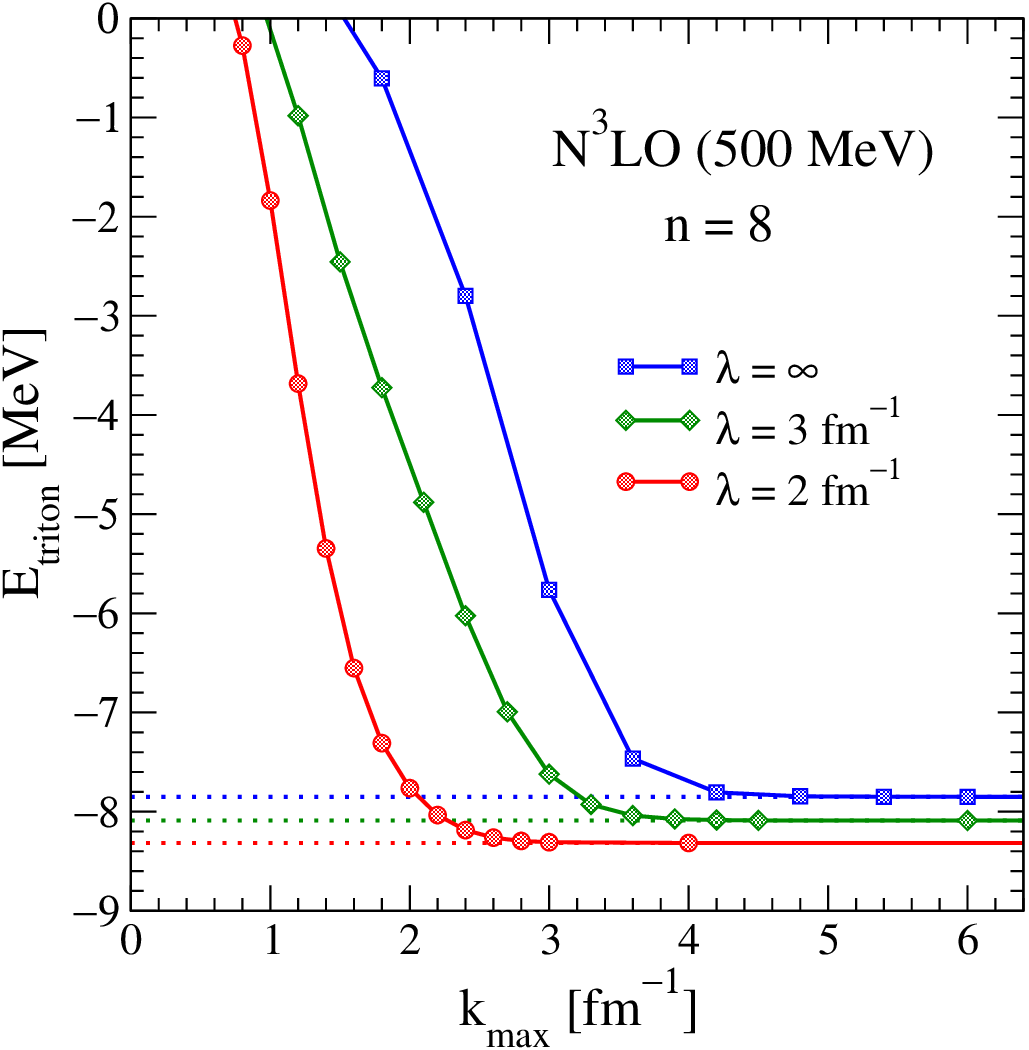}
 \vspace*{-.1in}
 \caption[]{Left: Decoupling in the $^1$S$_0$ phase shift for the
  Argonne $v_{18}$ NN potential~\cite{Bogner:2007srg}.  Right: Decoupling in the triton
  with the N$^3$LO chiral EFT potential of Entem and
  Machleidt~\cite{Bogner:2007rx}.
  }
  \label{fig:ps}
 \vspace*{-.1in}
\end{figure}

In the left panel of Fig.~\ref{fig:ps}, the $^1$S$_0$ phase shift
for the Argonne $v_{18}$ NN potential is shown up to 800\,MeV lab
energy.  The phase shifts for the SRG potential $V_s$ are indistinguishable
at any $\lambda$ because the evolution is exactly unitary at the two-body level.
To test decoupling, the original and evolved SRG potential
(to $\lambda = 2\,\mbox{fm}^{-1}$) are smoothly set to zero for momenta
above $k_{\rm max} = 2.2\,\mbox{fm}^{-1}$.  The SRG phases are unchanged
up to the corresponding $E_{\rm lab}$, so high momenta are \emph{not} needed. 
The AV18 phases are completely
changed because even low-energy observables have contributions from high momentum,
which has led to the misconception that  
high-energy phase shifts are important for nuclear
structure~\cite{Bogner:2007srg}.

\begin{figure}[hbt]
  \includegraphics*[width=2.7in]{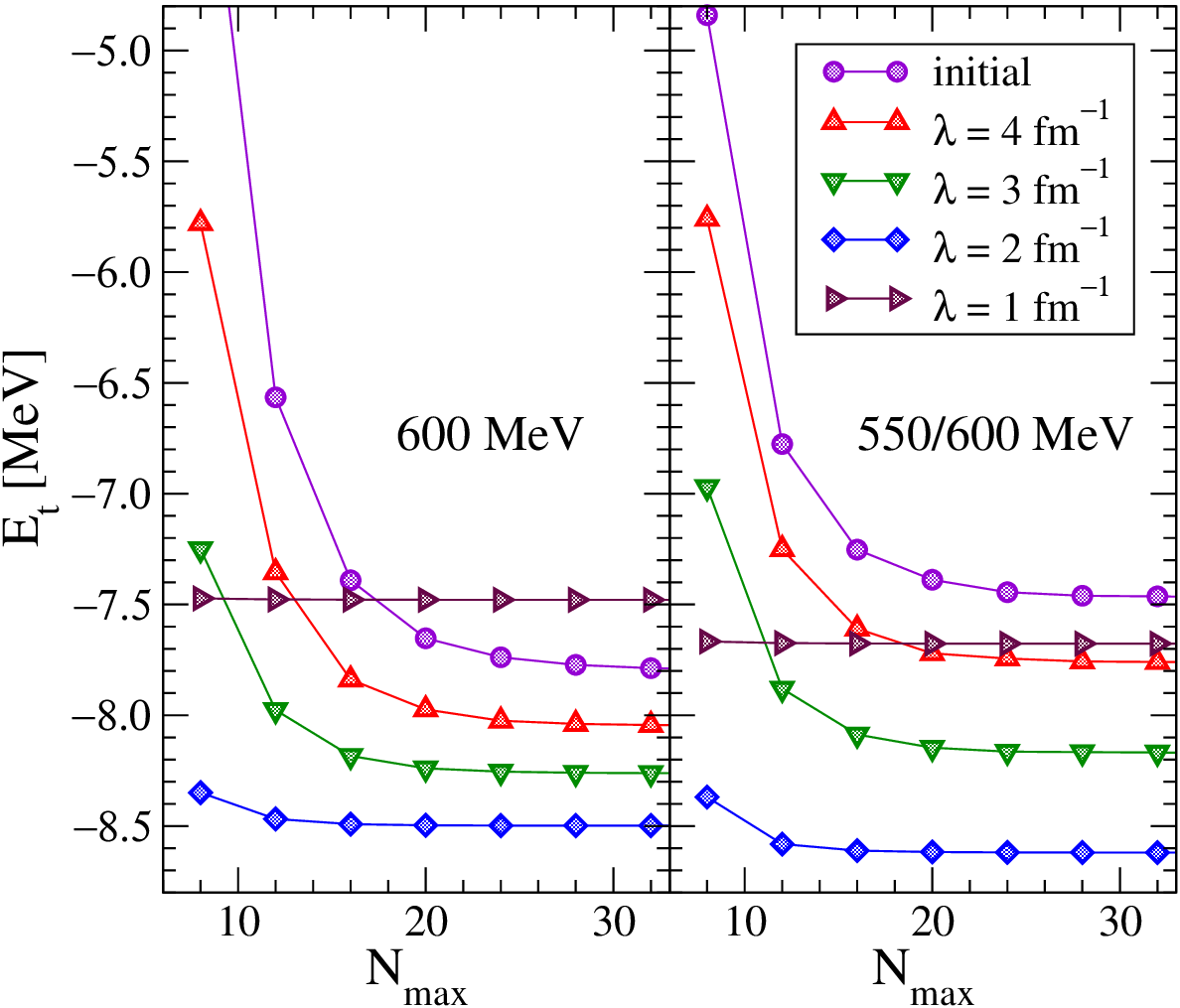} 
 \hfill
 \includegraphics*[width=2.5in]{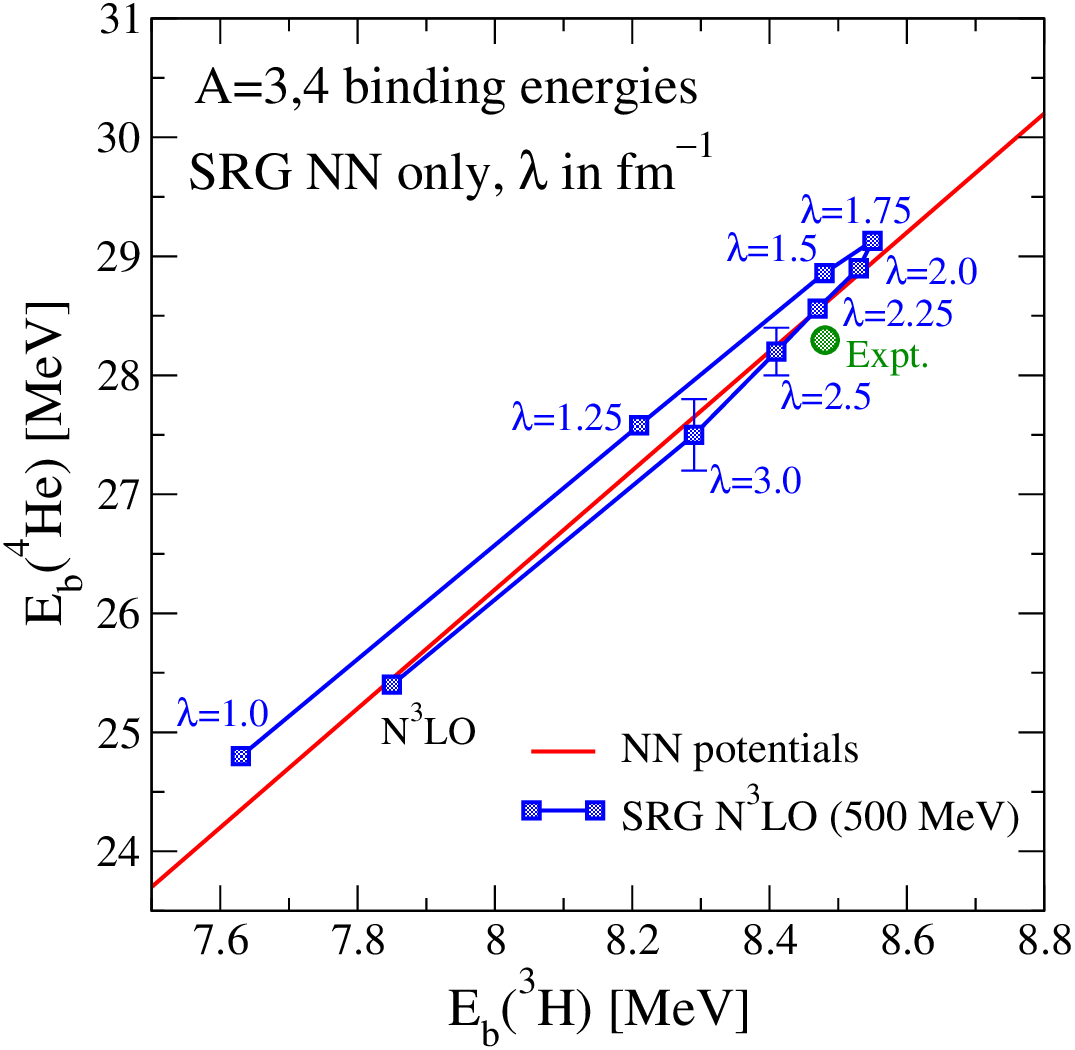}
 \vspace*{-.1in}
 \caption[]{Left: Convergence in the triton~\cite{Bogner:2007srg}.  
 Right: Tjon line traced out by SRG-evolved NN potentials,
 labeled by $\lambda$~\cite{Bogner:2007rx}.}
 \label{fig:triton}
 \vspace*{-.1in}
\end{figure}

A similar story for the triton ground-state energy with a chiral
EFT N$^3$LO potential is seen in the
right panel, which shows the energy as a function
of $k_{\rm max}$.  
Because of decoupling,
the full answer for smaller $\lambda$ is reached for lower $k_{\rm max}$.   
A consequence is faster convergence in 
variational and
\textit{ab initio}
few-body calculations, as shown in the left panel of Fig.~\ref{fig:triton}
for two N$^3$LO potentials, where the energy is plotted against the size
of a harmonic oscillator basis.
A complete study with NN potentials in the no-core shell model
is given in ref.~\cite{Bogner:2007rx}.

\begin{figure}[hbt]
  \includegraphics*[width=2.6in]{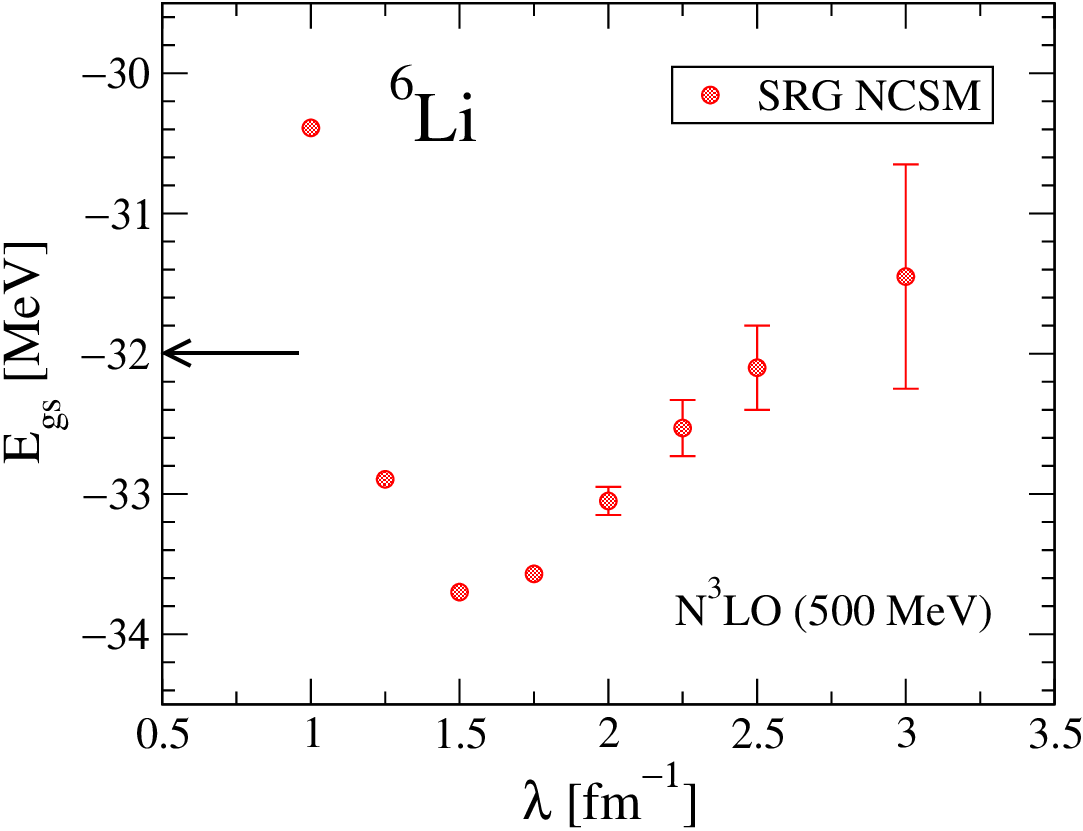} 
 \hfill
 \includegraphics*[width=2.5in]{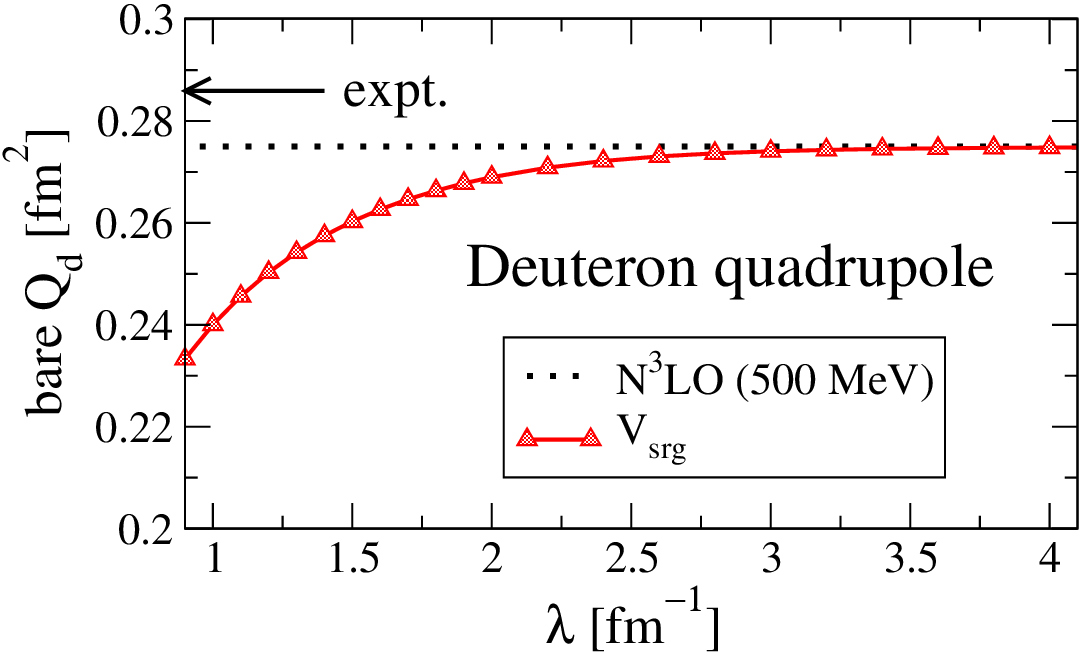}
 \vspace*{-.1in}
 \caption[]{Left: Ground-state energy 
 in $^6$Li vs.\ $\lambda$~\cite{Bogner:2007rx}.  
 Right: Running of the bare (unevolved) deuteron
  quadrupole moment.}
 \label{fig:fewbody}
 \vspace*{-.1in}
\end{figure}

The commutators in Eq.~(\ref{eq:Hflow}) imply that the evolving Hamiltonian
will have many-body interactions to all orders (i.e., insert second-quantized
operators).
Thus there will always be a truncation and the evolution will only be
approximately unitary.
The present calculations evolve only the
NN part, which explains why different converged triton energies are
seen in Fig.~\ref{fig:triton}.
This is a controlled approximation in the range of $\lambda$ for which
the variation is comparable to the truncation error inherent in
the initial EFT Hamiltonian.     The variation is seen to be natural
in Fig.~\ref{fig:triton} and the left panel of
Fig.~\ref{fig:fewbody}, which also shows the improved convergence
(decreasing error bars) for smaller $\lambda$~\cite{Bogner:2007rx}.

Including the 3N interaction is
essential for nuclear structure.
The SRG evolution only modifies the short-distance part of the potential
or operators.  This is illustrated by the weak running of the bare quadrupole
moment in the right panel of Fig.~\ref{fig:fewbody}.  
Since the chiral EFT 3N force
will be modified only at short distance, 
a good first approximation should be to simply
re-fit its two short-distance coefficients at each $\lambda$.
In parallel, we are implementing the 3N evolution 
by applying Eq.~(\ref{eq:Hflow}) in the three-particle space,
which does not require solving the full three-nucleon
problem~\cite{Bogner:2006srg}.
The evolution of the NN interaction is independent of spectators and the
equation for the 3N interaction has no disconnected pieces.
A model calculation that introduces a diagrammatic treatment
is described in ref.~\cite{Bogner:2007qb}.

\begin{acknowledge}
I gratefully acknowledge my collaborators
E. Anderson, S. Bogner, E. Jurgenson, P.~Maris, R. Perry, A. Schwenk,
and J.~Vary.
This work was supported in part by the National Science Foundation
under Grant Grant Nos.~PHY--0354916 and PHY--0653312, 
and the UNEDF SciDAC Collaboration under DOE Grant 
DE-FC02-07ER41457.
\end{acknowledge}

\end{document}